# Radiative Pattern of Intralayer and Interlayer Excitons in Two-Dimensional WS₂/WSe₂ Heterostructure

Mohammed Adel Aly[1,2,+], Manan Shah[1,+], Lorenz Maximilian Schneider[1,+], Kyungnam Kang[3], Martin Koch[1], Eui-Hyeok Yang[3,*], and Arash Rahimi-Iman[1,4,*]

[1]*Faculty of Physics and Materials Sciences Center, Philipps-Universität Marburg, Marburg, 35032, Germany*

[2]*Physics Department, Faculty of Science, Ain Shams University, 11566 Cairo, Egypt*

[3]*Department of Mechanical Engineering, Stevens Institute of Technology, Hoboken, NJ 07030, USA*

[4]*1. Physikalisches Institut, Justus-Liebig-Universität Gießen, Gießen, 35390, Germany*

*+These authors have equally contributed to this work*

*\*Corresponding authors*

**Abstract:**

**Two-dimensional (2D) heterostructures (HS) formed by transition-metal dichalcogenide (TMDC) monolayers offer a unique platform for the study of intralayer and interlayer excitons as well as moiré-pattern-induced features. Particularly, the dipolar charge-transfer exciton comprising an electron and a hole, which are confined to separate layers of 2D semiconductors and Coulomb-bound across the heterojunction interface, has drawn considerable attention in the research community. On the one hand, it bears significance for optoelectronic devices, e.g. in terms of charge carrier extraction from photovoltaic devices. On the other hand, its spatially indirect nature and correspondingly high longevity among excitons as well as its out-of-plane dipole orientation render it attractive for excitonic Bose-Einstein condensation studies, which address collective coherence effects, and for photonic integration schemes with TMDCs. Here, we demonstrate the interlayer excitons' out-of-plane dipole orientation through angle-resolved spectroscopy of the HS photoluminescence at cryogenic temperatures, employing a tungsten-based TMDC HS. Within the measurable light cone, the directly-obtained radiation profile of this species clearly resembles that of an in-plane emitter which deviates from that of the intralayer bright excitons as well as the other excitonic HS features recently attributed to artificial superlattices formed by moiré patterns.**

## Introduction

During the last decade, atomically thin two-dimensional (2D) materials based on transition metal dichalcogenides (TMDC) have attracted immense interest due to their extraordinary light–exciton interaction[1]. Owing to their wide tunability of the band gap across the visible spectral range[2] and virtually lattice-matching free incorporation into existing device technology platforms[3], they gained popularity for nanoscale optoelectronics[4,5], as well as quantum technologies and photonic integration schemes[6–9], to name but a few examples. The large electronic energy gaps define the optical properties of 2D TMDC monolayers (MLs), which are governed by neutral and charged excitonic species with extraordinarily large binding energy for (Wannier–Mott type) crystal excitons due to their strong Coulomb interaction, reduced dielectric screening and quantum confinement[10–12]. Furthermore, additional excitonic states beyond charged excitons (trions)[13], such as the biexcitons[14] as well as other complex states[15,16], also affect the dynamics and spectral features[17–20] of the monolayers, particularly for high-quality samples and predominantly at cryogenic temperatures[21–24].

Vertical van-der-Waals (vdW) heterostructures (HSs)[3,25], which can be straight-forwardly assembled from TMDC, hBN and graphene, offer exciting opportunities to study exciton physics, as well as novel and extraordinary phases of correlated matter[26]. These features are understood to play a crucial role in developing next-generation (integrated) photonic and (nanoscale) optoelectronic devices with the help of such artificially stacked multilayer-configured ML/few-layer systems. With the observation of superconductivity in bilayer graphene[27–29] and other exotic quantum states[30–33] including bands with topological properties [34], the interlayer twist angle even gave the whole field a new twist, as different investigations have shown recently[35–37]. HSs comprising ML TMDCs with type-II band alignment are particularly attractive for the effective formation of charge-transfer excitons (interlayer excitons)[38] and consecutive efficient dissociation (charge separation towards electronic contacts) in the heterobilayer (HBL) interface system, benefiting from band hybridizations and electronic band edge offsets. These properties are essential in the context of ultrafast photodetection with HS devices and an efficient photovoltaic effect. Moreover, interlayer excitons are strongly bound ($E_{\text{bind.}} > 100$ meV) and typically persevere at elevated temperatures up to room temperature and exhibit considerably longer lifetimes (in the nanosecond scale) than their intralayer counterparts (e.g., neutral ML excitons). Furthermore, they offer the possibility of forming Bose-Einstein-like condensate coherent states[26] through spontaneous coherence formation below a critical temperature for these dipolar out-of-plane excitons, the demonstration of which requires elaborate investigations of temporal and spatial coherence properties of the quantum-degenerate bosonic state.

In the type-II HBLs established by arbitrarily stacked TMDC monolayers (MLs), the interlayer exciton can be formed between electrons and holes present in two different MLs, which come with a layer-to-layer twist-angle degree of freedom that affects their energetics and dynamics[39]. Remarkably, this interlayer state had been well identified both in photoluminescence (PL) and reflection contrast (RC)[40–42]. Furthermore, due to lattice mismatch and twist angles, artificial periodic potentials are induced with a direct impact on the excitonic phase in the TMDC HBLs – by the so-called moiré superlattice[32,43,44]. This moiré-pattern induced periodic potential modulation, which is tunable in terms of supercell length scale by the twist angle, provides a powerful tool, for instance, in configuring quantum phenomena in 2D HSs[27], such as by tuning the HBL's electronic structure in and out of flat band situations or tailoring lateral trapping potentials and interaction strengths.

The aforementioned spatial separation between Coulomb-bound electrons and holes (i.e., the spatially indirect nature of excitons) is understood to result in a considerable out-of-plane dipole contribution which differs from the in-plane dipoles within the ML (i.e., the intralayer excitons). Recently, far-field studies were carried out on different TMDC MLs[20,45], on a single-crystal layered perovskite[46] and on a 3D crystal such as InSe[47] to investigate their radiative patterns, i.e., the different out-of-plane excitons' emission profiles. Driven in part by the distinct excitonic properties of interlayer excitons (X$_{IL}$) and also the fact that there are no such experiments, which focus on clearly disentangling the radiation patterns for different interlayer and intralayer excitons in TMDC heterosystems, a sound understanding and verification of the emission profile for interlayer excitons is sought.

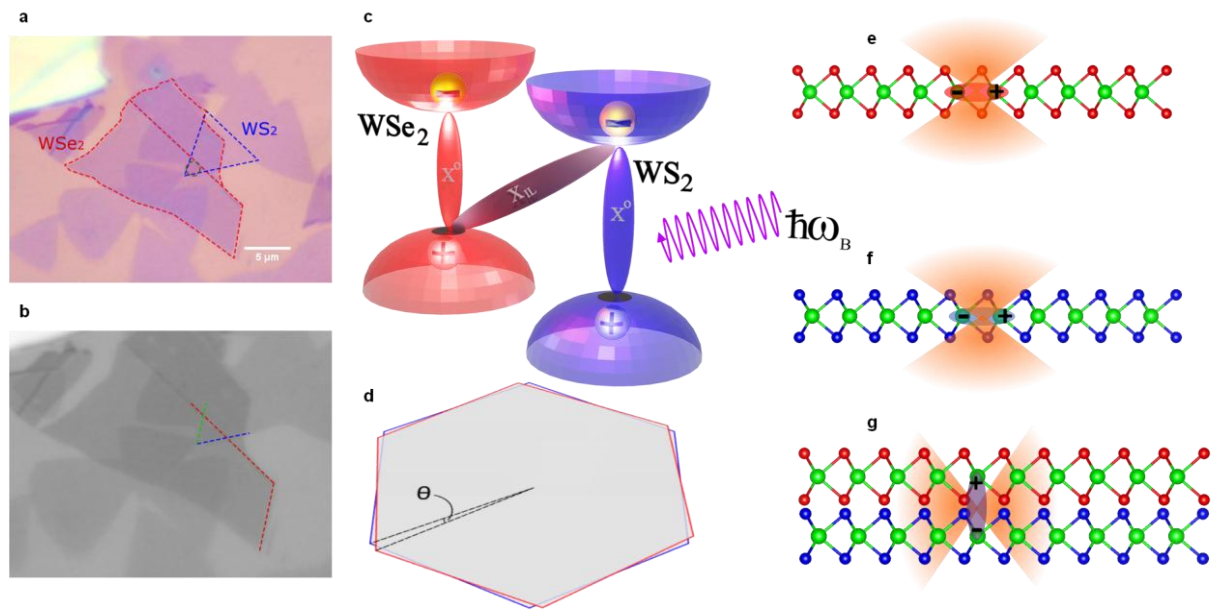

**Figure 1. Representation of different excitonic features (transitions) and their corresponding radiative emission patterns of WS$_2$/WSe$_2$ vdW heterostructure (HS).** (**a-b**) Optical micrograph for WS$_2$/WSe$_2$ heterobilayer (HBL) on SiO$_2$/Si$_2$ substrate (300 nm thermal oxide layer) taken under 100x objective. (**a**) White light color readout. The relevant monolayer (ML) area for WSe$_2$ is indicated by the red-dotted frame lines, for WS$_2$ by the blue-dotted ones. The dotted circle represents the HBL spot under investigation. (**b**) Intensity profile of the green channel. The red-line marked edges of the top flake and blue/green-lined edges of the CVD (bottom) flake of the HBL region indicate the twist angle θ, here estimated visually to be 56°. (**c**) Schematic drawing of type-II band alignment of the HS. The A-exciton (X$^0$) transitions are sketched in red and blue colors for WSe$_2$ and WS$_2$, respectively. Furthermore, the charge-transfer interlayer exciton (ILX, X$_{IL}$) is indicated as a gray colored transition. (**d**) A sketch of two twisted lattice Brillouin zones in *k*-space is shown, hinting at the possible phase-space mismatch at the corners (representing K and K' valleys of TMDCs) with increasing angle of θ. (**e-g**) In WS$_2$ and WSe$_2$ MLs, an in-plane dipole orientation for intralayer excitons and their respective out-of-plane radiative emission patterns can be seen. For interlayer excitons, which are formed across the interface of two different HS layers, an out-of-plane dipole orientation and the in-plane radiation pattern is indicated.

In this study, we provide a direct measurement of the excitonic luminescence for different states (with in-plane and out-of-plane dipole orientations) of our model-type Tungsten-based HBL system, which can be addressed by our angle-resolved photoluminescence technique (ARPL). Moreover, our analysis supported by numerical modeling shows that the PL signatures for WS$_2$ and WSe$_2$ MLs indeed originate exclusively from in-plane dipoles. In contrast, we unravel with our experiment the out-of-plane nature of charge-transfer exciton signatures obtained by PL at 10 K, which is in agreement with predictions and our modeling data for these interlayer quasi-particles formed across the HBL interface in the WS$_2$/WSe$_2$ ML–ML HS. Here, clear evidence is given that this radiative pattern of HBL charge-transfer excitons is distinct from that of the previously-found moiré-induced states in this system, the emission profile of which resembles the characteristics of intralayer (ML) excitons, as well as from that of emission from localized states in WSe$_2$. This provides an unrivalled means of differentiation between these species present in HBLs.

## Experiment

In this work, a WS$_2$/WSe$_2$ HS is studied. The WS$_2$ ML was grown by the CVD technique, whereas the WSe$_2$ ML was exfoliated from bulk crystal, then stacked over WS$_2$ by the dry-stamping technique in order to assemble a HS with a hybrid-production-type HBL region. In **Fig. 1**, an optical micrograph of the home-built HS is shown as a full-color CMOS-camera image (**a**) and the gray-scale image of the green channel (**b**). Red and blue channel micrographs are available in the **Supporting Information (Fig. SI.2)**. ML regions and the spot of interest are indicated by dotted lines and labels in the color micrograph, whereas the twist angle can be extracted from crystal flake edge orientations marked in the gray-scale micrograph. Here, the visually-extracted angle amounts to 56° between the lattices. As previously reported[40], WS$_2$/WSe$_2$ vdW HSs deliver a type-II band alignment between the constituting MLs. Therefore, the electrons from the WSe$_2$ conduction band (CB) can efficiently be injected into the WS$_2$ CB, provided that the lattice-twist induced phase mismatch at the respective valleys across the interface is not too significant to impede this process – not specifically looking at the impact on the moiré patterns and their implications on the electronic hybridization at this point.

Accordingly, the system ends up with spatially-indirect ILX across the HBL interface with the constituting electron and hole residing in different layers. Thus, the HS exhibits a lower effective optical band gap indicated with the transition for ILX in **Fig. 1c**. Furthermore, the expected radiation pattern for different intralayer excitonic species (A-exciton in WS$_2$ and WSe$_2$) and the ILX of this HBL are sketched in **Fig. 1e-g**. The out-of-plane emission pattern arising from the in-plane dipoles (A-excitons) and the corresponding in-plane pattern associated with the out-of-plane dipoles (ILX) are visualized schematically.

ARPL measurements were performed under nonresonant 2.3-eV continuous-wave (CW) laser excitation (532-nm frequency-doubled Nd:YAG) with a spot size of 2–3 µm to examine the radiative nature of different dipole species. The sketched experimental setup[48] is presented in the **Supporting Information (Figure SI.1)**.

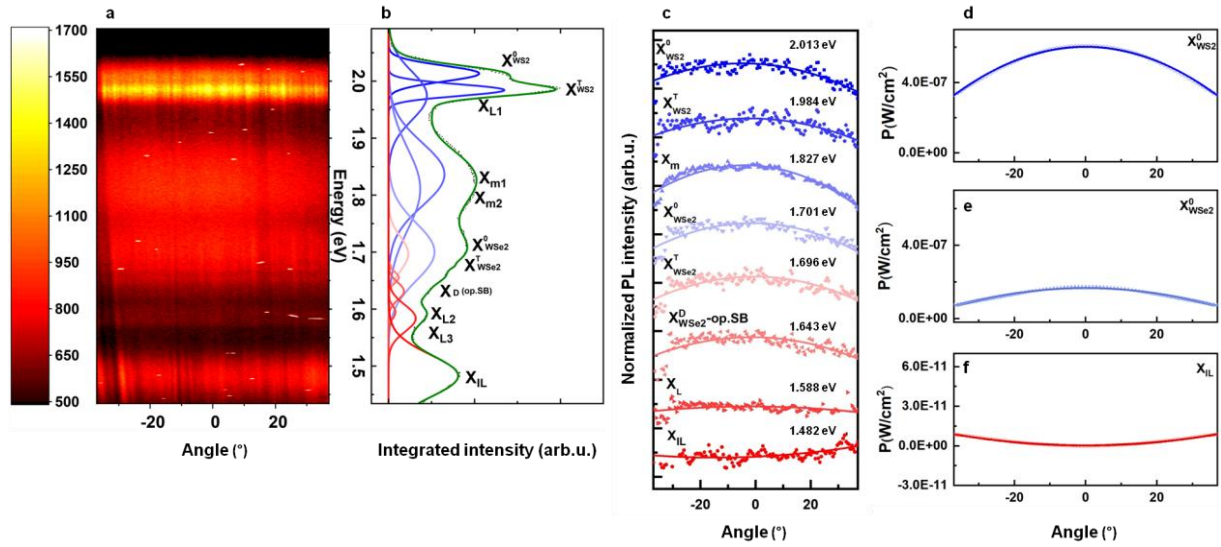

**Figure 2. Angle-resolved micro-photoluminescence (ARPL) and their emission pattern**. (**a**) ARPL emission spectrum in 2D contour-plot style (linear intensity map) for the WS$_2$/WSe$_2$ HS at 10 K under nonresonant CW laser excitation. The corresponding angle-integrated line spectrum is depicted to the right (**b**). Recorded data (points) are shown with multi-Gaussian-fit model curves (solid lines, color-coded from blue to red with decreasing mode energy). (**c**) Extracted radiation emission patterns (color-coded data points) for different excitonic species seen in (**a**) and their intensity variation with emission angle. (**d-f**) Simulated farfield emission (data points) at three different spectral positions, given in P (W/m$^2$), for hypothetical in- and out-of-plane dipole emitters (X$^0$ at 2.013 eV, 1.70 eV, and X$_{IL}$ at 1.482 eV, respectively) in the simulated HS environment. Solid lines in (**c-f**) indicate the differently-strong parabolic trends towards higher angles, as guides to the eyes.

The spectrally-resolved back-focal-plane (*far-field*: FF) PL emission[23,24] of the different excitonic species is displayed in the contour-diagram intensity spectrum of **Fig. 2a**. The linear profile scales from dark (low) to white (high counts). Next to the 2D plot, the angle-integrated emission is summarized as labeled line-spectrum with multi-peak Gaussian model as fit curves (**Fig. 2b**). The color code is reused throughout the figure with blueish and reddish colors representing the higher- and lower-energy features, respectively. According to the inherited convention from previous works, the different excitonic species are labeled at their corresponding energies[20,40,49,50]. In **Fig. 2c**, the displayed radiation profiles are obtained from spectral integration over their relevant particular energy region (see **Table 1**). Guides to the eyes indicate the curvature. While basically all species feature a convex shape, i.e., the maximum output centered at normal incidence, the ILX profile reveals an inverted behavior with a concave shape, i.e. the minimum being centered. Simulation data augments the analysis and reproduces these profiles as displayed for the representative dipole species with distinct orientations in **Fig. 2d-f**. This analysis is performed in a similar fashion as for the bright- and dark-exciton study by *Schneider et al.*[20] In the following, our findings for HBL features are examined and summarized. Additional HS and ML characterization by PL, reflection-contrast and Raman spectroscopy is briefly summarized in the **Supporting Information (Fig. SI.4–7)**.

## Results and Discussion

In order to extract and analyze different excitonic features, the ARPL spectrum was integrated over the full angle range. The corresponding line-spectrum is plotted in **Fig. 2b** next to the angle-resolving 2D contour plot in (**a**). Here, the spectrum indicates that the WSe$_2$ signal is quenched in the HBL region's emission, and that the interlayer signal is enhanced due to the expected charge transfer (sketched in **Fig. 1c**). For clarity, **Table 1** lists the parameters obtained from Gaussian fitting analysis applied on this line-spectrum. The extracted excitonic features and their energetics in our HBL's PL agree with those reported by *Yu et al.*[38] and *Shi et al.*[51] for X$^0$, X$^T$ in WS$_2$ and the HBL's X$_{IL}$. Whereas, X$_0$, X$_T$ in the ML WSe$_2$ region and the HBL's moiré-attributed feature X$_m$ agrees well with the data reported for the fully exfoliation-based stack by *Shah et al.*[40] Moreover, dark excitons X$^D$ match those previously reported by *Schneider et al.*[20]

From **Fig. 2**, one can distinguish the interlayer signal from the other excitonic features due to its lower emission energy at 1.482 eV. Furthermore, the difference in emission patterns extracted from **Fig. 2a** can be compared. **Figure 2c** provides an overview of different radiation patterns for all clearly spectrally-distinguishable excitonic features. Here, the interlayer exciton and other excitonic features are discriminated by their concave and convex intensity distribution along the emission-angle axis, respectively. This behavior is directly attributed to an in-plane emission behavior of the interlayer exciton (out-of-plane dipole orientation) and the emission orientation perpendicular to the 2D-material plane of other excitonic features (intralayer dipoles), respectively, e.g., A-exciton and -trion for the two involved MLs. It is apparent from this graph that most of the excitonic features show a similar behavior as neutral excitons. Markedly, on the other hand, the moiré feature emission is most directional and exhibits the most pronounced convex intensity profile. This can be attributed to the collective emission from the artificial quantum-dots/nanoislands array forming a single wavefront with out-of-plane propagation. Such moiré-induced array defines the 2D-periodic in-plane potential modulation and traps moiré excitons within nanosize supercells, delivering a plethora of in-plane emitters acting as arrayed nanoantennae. In contrast to in-plane and out-of-plane oriented species, PL lines from defect states in the TMDCs, such as the X$_L$ signal, expectedly exhibit no observable angle dependencies compared to the other distinct HS exciton types (cf. flat X$_L$ intensity profile in **Fig. 2c**), according to the statistical emission distribution from localized emitters.

**Table 1. Excitonic features of the HBL system extracted from the angle-integrated PL line-spectrum.** Excitonic mode energies and their corresponding linewidths (full-width at half-maximum) are obtained through multi-peak Gaussian-curve fitting. The different excitonic species are labeled as shown in **Fig. 2**.

| Excitonic feature | X$^0_{WS2}$ | X$^T_{WS2}$ | X$_{L1}$ | X$_{m1}$ | X$_{m2}$ | X$^0_{WSe2}$ | X$^T_{WSe2}$ | X$^D$ WSe2$^-$ op.SB | X$^D$ WSe2$^-$ op.SB | X$_{L2}$ | X$_{L2}$ | X$_{IL}$ |
|---|---|---|---|---|---|---|---|---|---|---|---|---|
| Energy (eV) | 2.013 | 1.984 | 1.974 | 1.836 | 1.810 | 1.701 | 1.696 | 1.655 | 1.631 | 1.594 | 1.583 | 1.482 |
| Linewidth (eV) | 0.033 | 0.0244 | 0.100 | 0.130 | 0.167 | 0.118 | 0.070 | 0.023 | 0.035 | 0.026 | 0.055 | 0.087 |

To support our experimental findings, an electromagnetic simulation was carried out for two differently-oriented emitter types (in-plane and out-of-plane dipoles) to calculate their far-field (FF) pattern. Simulated FF patterns, i.e. the angle-dependent irradiance (W/cm²) for a given dipole in the (dielectric/semiconducting) host environment at predefined emission energy, are displayed in **Fig. 2d-f** in a linear intensity scale: The radiation profile for the A-exciton of WS$_2$ (**d**) and of WSe$_2$ (**e**) clearly show the expected maximum emission at normal incidence with a inverse-parabola-like decay to the sides; whereas, the interlayer exciton X$_{IL}$ (**f**) exhibits an opposite behavior for the same angle range and drastically lower outcoupling from the layered HBL host environment. Thereby, it is clearly demonstrated that the simulated radiation profiles are in a good agreement with the experimental results obtained from the angle-resolved measurements.

This experiment as well as analysis confirms that the obtained emission patterns for the interlayer exciton and the other excitonic features arise from intrinsic out-of-plane and in-plane dipole orientation, respectively, of the Coulomb-bound quasiparticles of the HBL region under investigation.

**Conclusion**

In summary, we have analyzed the photoluminescence emission pattern for WS$_2$/WSe$_2$ van der Waals semiconductor heterobilayers by means of angle-resolved spectroscopy. Our ARPL measurements enable direct access to the emission behavior of different excitonic modes with different dipole orientation in 2D TMDC HBLs and constituent MLs. Our findings highlight the markedly in-plane radiation profile for the interlayer excitons formed from spatially-separated electron–hole pairs across the HS interface, which are distinct in their behavior compared to the other excitonic modes. It is demonstrated at cryogenic temperatures that most of the excitonic modes show a convex angle-dependent intensity profile while the interlayer exciton's signature is a concave profile. Such information enables their utilization in different photonics applications. This study motivates further

*k*-space resolved measurements analysis for interlayer excitons in various type-II heterojunction systems. Moreover, further investigations of angle-resolved HS emission involving external bias across the heterojunction and gate tuning effects are envisioned.

## Methods

**Sample fabrication.**

The tungsten disulfide WS$_2$ MLs studied in this work were grown via Low-Pressure Chemical Vapor Deposition (LPCVD) onto a clean substrate (~300nm SiO$_2$ on Si).

5 nm thick WO$_3$ was evaporated from pellets onto a source substrate sandwiched with a clean second substrate for growth, with no space. The sandwiched sample was loaded into the middle of a 3" quartz tube. For sulfur, we adopted the commonly published method for MoS$_2$ and WS$_2$ growth of placing solid sulfur powder in the furnace tube upstream of the growth area. A mechanical pump purged out the ambient gas to the base pressure of 850 mTorr. As the furnace was ramped in temperature at 15 ˚C/min, the reaction reduced WO$_3$ by hydrogen and subsequent sulfurization of the WO$_3$. The growth temperature was 900˚C. Ar gas was introduced from 150 ˚C to reduce moisture, and ambient gas and H$_2$ gas were supplied from 650˚C (increasing temperature) to 700˚C (decreasing temperature). The deposition pressure depends on the gas type and amount of flow rate. The best result was obtained at 4.5 Torr deposition pressure under 60 sccm H$_2$ flow rate. The reduction and sulfurization reactions require a higher temperature than the sulfur evaporation. By placing the sulfur at different places outside of the central heating area of the furnace, it evaporated at different times relative to the substrate temperature. At the optimized location for our furnace setup, the sulfur powder started to evaporate at 830˚C furnace temperature, and all sulfur powder was used up after about 30 minutes.

Tungsten diselenide (WSe$_2$) MLs were exfoliated from a commercial bulk crystal (*2D semiconductors Inc*) by using the common micro-mechanical-exfoliation and dry-stamping technique. The exfoliated flakes were transferred onto polydimethylsiloxane (PDMS). MLs were selected by optical contrast among the other flakes of different contrast and thickness using an optical microscope. Afterwards, the selected ML was transferred on top of a suitable CVD-grown WS$_2$ ML region with the viscoelastic stamp briefly heated to 90$^o$C for flake release. Thereafter, the PDMS was slightly lifted and the released WSe$_2$ ML remained attached to the target. The fabricated HS was then annealed at 300$^o$C for 4 hours under vacuum (~10$^{-6}$mbar) to enhance coupling between the layers.

**μ-PL measurement.**

Back-focal-plane imaging (i.e., Fourier-space imaging) was performed using a home-built 4-*f* μ-PL confocal optical microscope setup sketched in **Figure SI.1**. The Fourier-space spectroscopy technique used in previous works for ML signal characterization[20,23,24] allows us to measure the far-field PL signal, that is the intensity as a function of the emission angle (**Figure SI.3**). Experiments were performed using the 532-nm CW laser as an excitation source with an average power of 1 mW (corresponding to about 31 kW cm$^{-2}$) focused on the sample with a 40x (NA 0.6) microscope objective to a Gaussian spot with a diameter of approximately 2 μm. The main excitonic features and relevant signatures of the examined HBL are preserved over a wide irradiance variation range, while no degradation is observed at such excitation powers when remaining below the typical damage threshold for such samples. The maximum detectable angle corresponds to ±37°. The sample was mounted in a continuous-flow cryostat at a high vacuum (~10$^{-7}$ mbar) and was cooled down to 10 K. The PL emission from the sample was collected by the same objective and directed to the detection optics part. For analysis, a 550-nm long-pass filter was placed after the sample to suppress the laser light in the data acquisition part. The PL signal was mapped using a nitrogen-cooled charge-coupled device (CCD) Si camera attached to the imaging monochromator (*Princeton Instruments Acton SP2300*). By employing the full chip array of the CCD, the angle resolution (±1°) of the FF signal was obtained.

**Dipole emission simulations.**

By using CST simulation *microwave studio* packages, FF emission patterns for different dipole orientations were simulated. The simulation parameters were extracted from the literature. For WSe$_2$, the thickness was estimated to be 0.6 nm[20]. In order to simulate and calculate the emission profile, the Lorentz model was employed. The dispersion parameters for the Lorentz model were extracted from *Laturia et al* [52]. For WSe$_2$, the in-plane permittivity is $\epsilon_{(\infty,z)} = \epsilon_{(s,z)} = 7.5$. While for the out-of-plane permittivity, the following values were used $\epsilon_{(\infty,x,y)} = 15.1$, $\epsilon_{(s,xy)} = 15.3$, together with a damping frequency of 4.77 THz. Furthermore, for WS$_2$, the in-plane permittivity is $\epsilon_{(\infty,z)} = \epsilon_{(s,z)} = 6.3$, whereas for out-of-plane permittivity, the following values were used $\epsilon_{(\infty,x,y)} = 13.6$, $\epsilon_{(s,xy)} = 13.7$, together with a damping frequency of 4.77 THz. The contribution of the out-of-plane intrinsic oscillators (out-of-plane intraband excitons) to the permittivity is much weaker than that of the in-plane component, giving no significant contribution to the permittivity[20]. Therefore, this contribution was neglected for the simulation. For silicon and silicon oxide, values were extracted from the program library. The resulting FF patterns were analyzed by plotting cartesian plots of power flow at constant Azimuth without any means of polarization (absolute value).

**Visualization.**

The schematic depiction of the WS$_2$/WSe$_2$ HS and their constituent MLs in **Figure 1c** is based on crystallographic data provided by the *Materials Project*[53,54] and drawn by the *VESTA* software[55].

## Acknowledgements

The authors acknowledge financial support by the Deutsche Forschungsgemeinschaft (DFG: SFB 1083, RA 2841/5-1 and SPP 2244), by the Philipps-Universität Marburg and the Deutsche Akademische Austausch Dienst (DAAD). M.A.A. is grateful for support from the Egyptian Ministry of Higher Education and Scientific Research, as well as the DAAD. A.R.-I. thanks former team members D. Renaud and O. M. Abdulmunem for their help in the early stages of monolayer heterostructures experiments and setup constructions, respectively, as well as T. F. Heinz for initial discussions. Synthesis of WS$_2$ monolayers was supported in part by a National Science Foundation award (ECCS-1104870, and EEC-1138244). The authors also thank Siwei Chen and Shichen Fu for their assistance in growing WS$_2$. The authors thank G. Witte and D. Günder for assistance with the annealing of the samples and W. Heimbrodt for access to his commercial Raman microscope.

## Authors' contributions

A.R.-I. conceived the experiment and initiated the study on angle-dependent measurements in 2015. The joint work was guided by E.H.Y. and A.R.-I. Monolayer WS$_2$ synthesis was achieved by K.K. and E.H.Y. Heterostructure assembly as well as characterization was performed by M.A.A. with the support of M.S. and L.M.S. The experiment was established by L.M.S. and A.R.-I., and the structures measured by M.A.A. with the help of M.S., L.M.S. and A.R.-I. The results were interpreted, discussed and summarized in a manuscript by M.A.A., M.S. and A.R.-I. with the help of all coauthors.

## Corresponding author

Arash Rahimi-Iman: a.r-i@physik.uni-marburg.de

Eui-Hyeok Yang: eyang@stevens.edu

## Authors' statement/Competing interests.

The authors declare no conflict of interest.

## Data availability

The data that support the findings of this study are available from the corresponding author upon reasonable request.

## Additional information

Supplementary Information accompanies this paper.

## References

1. Wang, G. et al. Colloquium: Excitons in atomically thin transition metal dichalcogenides. Rev. Mod. Phys. 90, 021001 (2018).

2. Xia, F., Wang, H., Xiao, D., Dubey, M. & Ramasubramaniam, A. Two-dimensional material nanophotonics. Nat. Photonics 8, 899–907 (2014).

3. Novoselov, K. S., Mishchenko, A., Carvalho, A. & Castro Neto, A. H. 2D materials and van der Waals heterostructures. Science 353, aac9439 (2016).

4. Wang, Q. H., Kalantar-Zadeh, K., Kis, A., Coleman, J. N. & Strano, M. S. Electronics and optoelectronics of two-dimensional transition metal dichalcogenides. Nature Nanotechnology vol. 7 699–712 (2012).

5. Koppens, F. H. L. et al. Photodetectors based on graphene, other two-dimensional materials and hybrid systems. Nat. Nanotechnol. 9, 780–793 (2014).

6. Wei, G., Stanev, T. K., Czaplewski, D., Jung, I. W. & Stern, N. P. Interfacing monolayer MoS2 with silicon-nitride integrated photonics. in Integrated Photonics Research, Silicon and Nanophotonics, IPRSN 2015 371p (Optical Society of America, 2015). doi:10.1364/iprsn.2015.im4a.3.

7. Mak, K. F. & Shan, J. Photonics and optoelectronics of 2D semiconductor transition metal dichalcogenides. Nat. Photonics 10, 216 (2016).

8. Liu, X. & Hersam, M. C. 2D materials for quantum information science. Nature Reviews Materials vol. 4 669–684 (2019).

9. Peyskens, F., Chakraborty, C., Muneeb, M., Van Thourhout, D. & Englund, D. Integration of single photon emitters in 2D layered materials with a silicon nitride photonic chip. Nat. Commun. 10, 1–7 (2019).

10. He, K. et al. Tightly bound excitons in monolayer WSe2. Phys. Rev. Lett. 113, 26803 (2014).

11. Chernikov, A. et al. Exciton binding energy and nonhydrogenic Rydberg series in monolayer WS2. Phys. Rev. Lett. 113, 76802 (2014).

12. Ye, Z. et al. Probing excitonic dark states in single-layer tungsten disulphide. Nature 513, 214–218 (2014).

13. Plechinger, G. et al. Trion fine structure and coupled spin-valley dynamics in monolayer tungsten disulfide. Nat. Commun. 7, 1–9 (2016).

14. You, Y. et al. Observation of biexcitons in monolayer WSe2. Nat. Phys. 11, 477–481 (2015).

15. Barbone, M. et al. Charge-tuneable biexciton complexes in monolayer WSe2. Nat. Commun. 9, 1–6 (2018).

16. Chen, S. Y., Goldstein, T., Taniguchi, T., Watanabe, K. & Yan, J. Coulomb-bound four- and five-

particle intervalley states in an atomically-thin semiconductor. Nat. Commun. 9, 1–8 (2018).

17. Danovich, M., Zólyomi, V., Fal'ko, V. I. & Aleiner, I. L. Auger recombination of dark excitons in WS2 and WSe2 monolayers. 2D Mater. 3, 035011 (2016).

18. Ruppert, C., Chernikov, A., Hill, H. M., Rigosi, A. F. & Heinz, T. F. The Role of Electronic and Phononic Excitation in the Optical Response of Monolayer WS2 after Ultrafast Excitation. Nano Lett. 17, 644–651 (2017).

19. Fu, J., Cruz, J. M. R. & Qu, F. Valley dynamics of different trion species in monolayer WSe2. Appl. Phys. Lett. 115, 082101 (2019).

20. Schneider, L. M. et al. Direct Measurement of the Radiative Pattern of Bright and Dark Excitons and Exciton Complexes in Encapsulated Tungsten Diselenide. Sci. Rep. 10, 8091 (2020).

21. Cadiz, F. et al. Excitonic linewidth approaching the homogeneous limit in MoS2-based van der Waals heterostructures. Phys. Rev. X 7, 021026 (2017).

22. Ajayi, O. A. et al. Approaching the intrinsic photoluminescence linewidth in transition metal dichalcogenide monolayers. 2D Mater. 4, 031011 (2017).

23. Schneider, L. M. et al. Shedding light on exciton's nature in monolayer quantum material by optical dispersion measurements. Opt. Express 27, 37131 (2019).

24. Schneider, L. M. et al. Optical dispersion of valley-hybridised coherent excitons with momentum-dependent valley polarisation in monolayer semiconductor. 2D Mater. 8, 015009 (2021).

25. Geim, A. K. & Grigorieva, I. V. Van der Waals heterostructures. Nature 499, 419–425 (2013).

26. Wang, Z. et al. Evidence of high-temperature exciton condensation in two-dimensional atomic double layers. Nature 574, 76–80 (2019).

27. Cao, Y. et al. Unconventional superconductivity in magic-angle graphene superlattices. Nature 556, 43–50 (2018).

28. Yankowitz, M. et al. Tuning superconductivity in twisted bilayer graphene. Science 363, 1059–1064 (2019).

29. Lu, X. et al. Superconductors, orbital magnets and correlated states in magic-angle bilayer graphene. Nature 574, 653–657 (2019).

30. Cao, Y. et al. Correlated insulator behaviour at half-filling in magic-angle graphene superlattices. Nat. 2018 5567699 556, 80–84 (2018).

31. Tang, Y. et al. Simulation of Hubbard model physics in WSe2/WS2 moiré superlattices. Nature 579, 353–358 (2020).

32. Regan, E. C. et al. Mott and generalized Wigner crystal states in WSe2/WS2 moiré superlattices. Nature 579, 359–363 (2020).

33.   Li, T. et al. Continuous Mott transition in semiconductor moiré superlattices. Nat. 2021
      5977876 597, 350–354 (2021).

34.   Zhang, Y., Devakul, T. & Fu, L. Spin-textured Chern bands in AB-stacked transition metal
      dichalcogenide bilayers. Proc. Natl. Acad. Sci. U. S. A. 118, (2021).

35.   Hunt, B. et al. Massive Dirac Fermions and Hofstadter Butterfly in a van der Waals
      Heterostructure. Science 340, 1427–1430 (2013).

36.   Ribeiro-Palau, R. et al. Twistable electronics with dynamically rotatable heterostructures.
      Science 361, 690–693 (2018).

37.   Xu, Y. et al. Correlated insulating states at fractional fillings of moiré superlattices. Nature 587,
      214–218 (2020).

38.   Yu, J. et al.  Observation of double indirect interlayer exciton in WSe 2 /WS 2 heterostructure .
      Opt. Express 28, 13260 (2020).

39.   Yuan, L. et al. Twist-angle-dependent interlayer exciton diffusion in WS2–WSe2
      heterobilayers. Nat. Mater. 19, 617–623 (2020).

40.   Shah, M., Schneider, L. M. & Rahimi-Iman, A. Observation of Intralayer and Interlayer Excitons
      in Monolayered WSe2/WS2 Heterostructure. Semiconductors 53, 2140–2146 (2019).

41.   Tongay, S. et al. Tuning interlayer coupling in large-area heterostructures with CVD-grown
      MoS2 and WS2 monolayers. Nano Lett. 14, 3185–3190 (2014).

42.   Wang, K. et al. Interlayer Coupling in Twisted WSe2/WS2 Bilayer Heterostructures Revealed
      by Optical Spectroscopy. ACS Nano 10, 6612–6622 (2016).

43.   Tartakovskii, A. Moiré or not. Nature Materials vol. 19 581–582 (2020).

44.   Jin, C. et al. Observation of moiré excitons in WSe2/WS2 heterostructure superlattices. Nature
      vol. 567 76–80 (2019).

45.   Wang, G. et al. In-Plane Propagation of Light in Transition Metal Dichalcogenide Monolayers:
      Optical Selection Rules. Phys. Rev. Lett. 119, 47401 (2017).

46.   Fieramosca, A. et al. Tunable Out-of-Plane Excitons in 2D Single-Crystal Perovskites. ACS
      Photonics 5, 4179–4185 (2018).

47.   Brotons-Gisbert, M. et al. Out-of-plane orientation of luminescent excitons in two-
      dimensional indium selenide. Nat. Commun. 10, 1–10 (2019).

48.   Lippert, S. et al. Influence of the substrate material on the optical properties of tungsten
      diselenide monolayers. 2D Mater. 4, 25045 (2017).

49.   Chen, S. Y., Goldstein, T., Taniguchi, T., Watanabe, K. & Yan, J. Coulomb-bound four- and five-
      particle intervalley states in an atomically-thin semiconductor. Nat. Commun. 9, 1–8 (2018).

50.   Brem, S. et al. Phonon-Assisted Photoluminescence from Indirect Excitons in Monolayers of
      Transition-Metal Dichalcogenides. Nano Lett. 20, 2849–2856 (2020).

51.     Shi, J. et al. Twisted-Angle-Dependent Optical Behaviors of Intralayer Excitons and Trions in WS2/WSe2 Heterostructure. ACS Photonics 6, 3082–3091 (2019).

52.     Laturia, A., Van de Put, M. L. & Vandenberghe, W. G. Dielectric properties of hexagonal boron nitride and transition metal dichalcogenides: from monolayer to bulk. npj 2D Mater. Appl. 2, 6 (2018).

53.     Persson, K. Materials Data on WSe2 (SG:194) by Materials Project. (2014) doi:10.17188/1192989.

54.     Persson, K. Materials Data on WS2 (SG:194) by Materials Project. (2016) doi:10.17188/1197614.

55.     Momma, K. & Izumi, F. VESTA 3 for three-dimensional visualization of crystal, volumetric and morphology data. urn:issn:0021-8898 44, 1272–1276 (2011).

# Supporting Information

## Radiative Pattern of Intralayer and Interlayer Excitons in Two-Dimensional WS$_2$/WSe$_2$ Heterostructure

Mohammed Adel Aly[1,2,+], Manan Shah[1,+], Lorenz Maximilian Schneider[1,+], Kyungnam Kang[3], Martin Koch[1], Eui-Hyeok Yang[3,*], and Arash Rahimi-Iman[1,4,*]

[1]*Faculty of Physics and Materials Sciences Center, Philipps-Universität Marburg, Marburg, 35032, Germany*

[2]*Physics Department, Faculty of Science, Ain Shams University, 11566 Cairo, Egypt*

[3]*Department of Mechanical Engineering, Stevens Institute of Technology, Hoboken, NJ 07030, USA*

[4]*1. Physikalisches Institut, Justus-Liebig-Universität Gießen, Gießen, 35390, Germany*

+*These authors have equally contributed to this work*

*\*Corresponding authors*

### S1. Schematic diagram of the angle-resolving experimental setup

The excitation laser is shone on the sample in a home-built conventional microscope setup [48]. Light is focused and collected with a 40x (NA 0.6) microscope objective (MO) mounted on a three-directional (3D) translation stage, employed for focussing as well as beam adjustments. Typical circular spot-sizes on the sample amount to approximately 2 µm in Gaussian beam diameter. The emitted/reflected signal is collected through the same MO. The backscattered laser light is blocked in the detection path by a long-pass filter. The setup is sketched in **Fig. SI.1**.

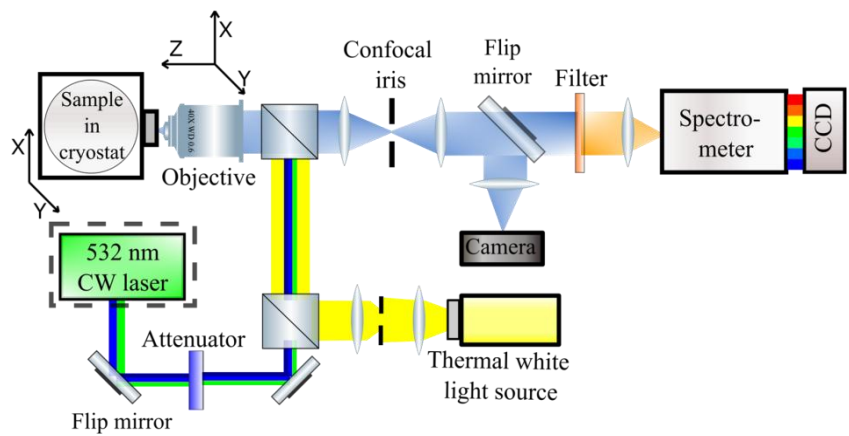

**Figure SI.1 | Schematic diagram of the micro-photoluminescence (PL) and reflection-contrast spectroscopy setup.**

For reflection-contrast measurements, the setup is used with the aid of an iris aperture in the near-field (NF) projection plane (sample's image plane) for spatial selection of the detection area in the order of 1 µm in diameter. The NF signal is further projected onto an imaging spectrometer's entrance slit and detected by an attached liquid-nitrogen cooled Si CCD camera. Moreover, the sample can be real-time monitored on the high-resolution CMOS (NF) video camera with the help of a flip mirror (**Fig. SI.2**).

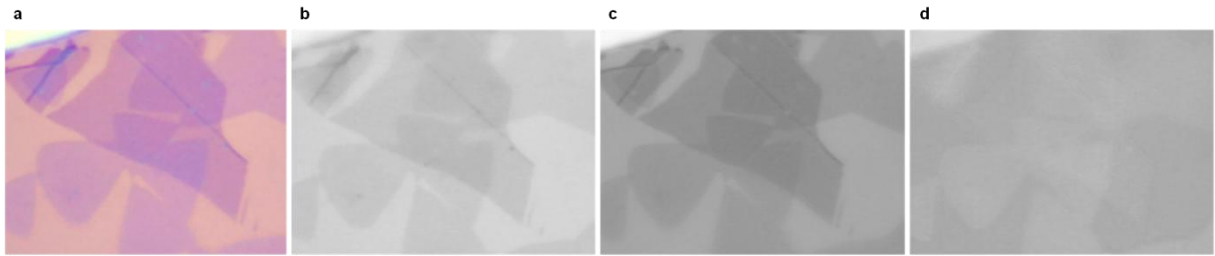

**Figure SI.2 | From left: Optical micrographs in full color and as gray-scale red, green as well as blue channel, respectively.**

For cryogenic measurements, a cryostat and liquid Helium flow can be used to control the temperature between 10 and 300 K. The cooled sample can be laterally translated under the microscope objective with the help of positioning stages.

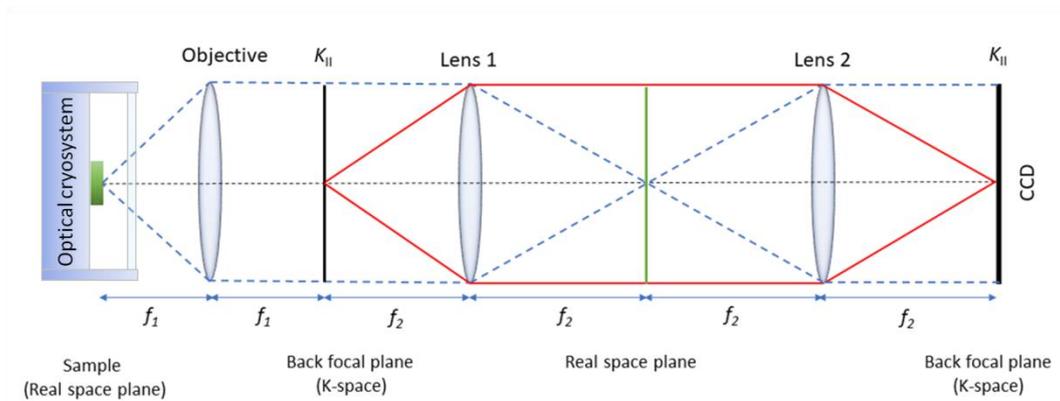

**Figure SI.3 | Sketch of the 4-*f* imaging configuration for angle-resolved spectroscopy.**

To obtain angle-resolved spectra similar to a previous experiment [20], the 4-*f*-projection configuration for Fourier-space imaging (**Fig. SI.3**) is applied which is capable of resolving the momentum space directly onto the imaging-monochromator CCD and previously used for dispersion characterization [23, 24]. The focal lengths of the optics are denoted by the parameter *f*.

## S2. Angle-integrated photoluminescence spectra

In **Fig. SI.4**, two distinct angle-integrated micro-PL (µPL) spectra for the heterobilayer (HBL) are presented, at room temperature (300 K, red) and cryogenic temperature (10 K, black), respectively. The PL spectra are recorded under 2.3-eV continuous-wave (CW) excitation, with imaging spectrometer CCD in line-spectrum mode. The linearly-scaled intensities are normalized and off-set for clarity (intensity in arb. u.). It is clearly seen that the CVD WS$_2$ A-exciton (1.955 eV) (also cf. [S7]) is red shifted compared to the A-exciton for hBN-supported exfoliated samples (~ 2.006 eV) reported before [S1, 40]. This is commonly attributed to strain for CVD samples which results from the growth of the monolayer (ML) in the furnace at elevated temperature. Furthermore, after growth and during cooling, due to the mismatch of thermal expansion coefficients by one order of magnitude between WS$_2$ monolayer and the SiO$_2$/Si substrate, strain/tension in the ML is imminent [S2]. In short, the WS$_2$ monolayer starts to shrink faster than the SiO$_2$/Si substrate during the cooling process as a consequence, but is stretched due to tension owing to surface adhesion between ML and substrate surface. Such conditions with impact on the band structure and vibrational modes can indeed affect exciton energies in PL spectra and are indicated by Raman mode shifts, respectively. Moreover, the decrease in the bandgap with increasing temperature in comparison to 10-K measurements is explainable through the enhanced electron–phonon interactions at elevated temperatures [S3], accompanied by changes in the bond lengths and evidenced through phonon-side bands as well as polaron formation [50, S4, S5, S6]. Due to the strong coupling between layers after annealing, the multi-exciton features occurring from hybridization of the electronic states of the original monolayer states can be observed and probed experimentally for the HBL region. Moreover, the formation of lower-energy interlayer excitons with their relatively bright emission centered at energies at about 1.48 eV (10 K) is obtained because of the strong overlap between the electron and hole wave functions in the targeted type-II heterosystem comprising ML TMDCs.

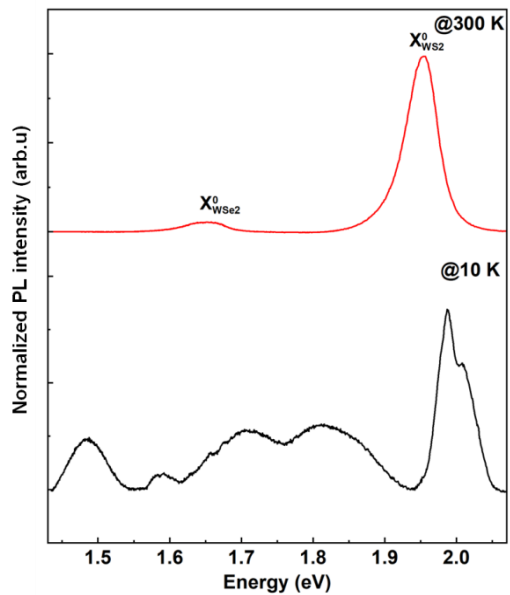

**Figure SI.4 | Angle-integrated μPL spectra for WS$_2$/WSe$_2$ HBL at 300 (red) and 10 K (black).**

## S3. Time-integrated reflection-contrast spectra

To access the properties of the excitonic features over a wide spectroscopic range, white-light reflection-contrast (RC) line-spectra are typically acquired. **Figure SI.5** shows the measured refection contrast (left axis) and its derivative with respect to energy (right axis) for the investigated WS$_2$/WSe$_2$ HBL at (**a**) 10 K and (**b**) 300 K. Minimum–maximum normalized RC data are linearly-scaled and offset in intensity. Different excitonic features are labelled for clarity and are in agreement with previous data for a fully-exfoliation-based HBL [40]. The spectra are obtained after annealing to complement aforementioned µPL data. At 300 K, the A-exciton for both WS$_2$ and WSe$_2$ are clearly visible, and much less pronounced the B-excitons identifiable spectrally in reflection-contrast derivatives. In contrast, at 10 K (**Fig. SI.5a**), considerably sharper lines are recorded, and more subtle resonances are detectable between the excitons for both individual ML materials. While for WSe$_2$ higher-order A-exciton states are indicated above 1.75 eV, its B-exciton mode obscures the clear detection of such states for the WS$_2$ A-exciton above 2.1 eV. Here, the broad ILX resonance below 1.5 eV in RC is washed out in the derivative chart, while signatures from possible moiré features between 1.75 and 1.95 eV may be hidden between the more clearly resolvable A-2s, 3s and higher states of WSe$_2$ in that spectral region. The small feature at 2.1 eV attributed to a B-trion of the WSe$_2$ layer overlaps with the high-energy flank of the WS$_2$ A-exciton, whereas the B-trion of the WS$_2$ layer coincides with the position where the B-trion is expected.

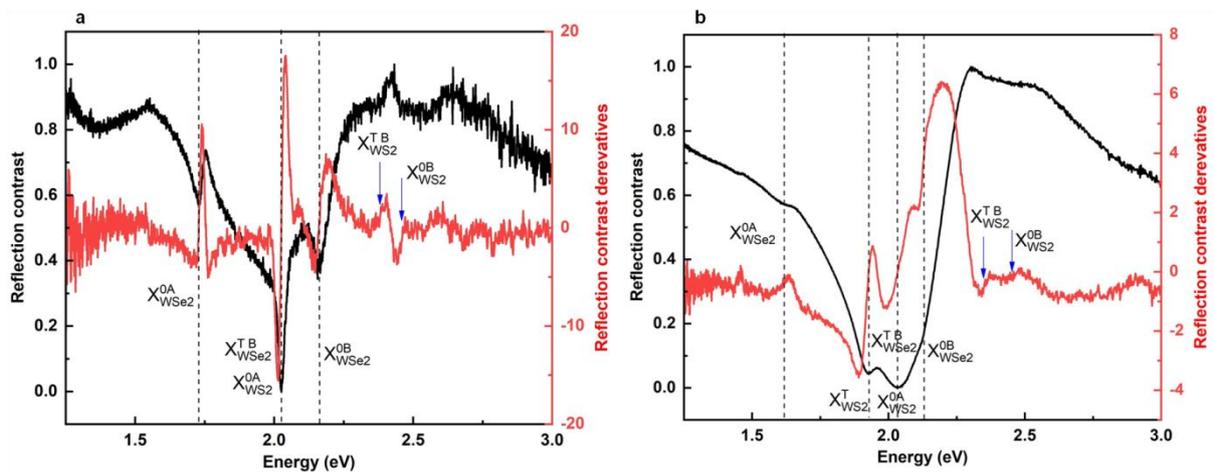

**Figure SI.5 | Time-integrated reflection-contrast data for WS₂/WSe₂ HBL region at (a) 10 and (b) 300 K**.

## S4. Raman signatures

For the detection of characteristic 2D TMDC Raman modes of the HBL system, a separate commercial Raman setup with a 100x objective (NA 0.9) with 514.8-nm Ar-ion laser as an excitation source is employed. The optical power is kept below 1 mW to avoid sample damage. The signal collected from the same objective is directed through the confocal microscope setup to a spectrometer with 1200-gr/mm grating and a liquid-nitrogen cooled Si CCD [S7]. **Figure SI.6a and b** show Raman modes of the HBL region recorded before and after annealing at 300°C, respectively. Gaussian-curve fits to the experimental data (points) are displayed (solid lines) and the features are labelled for clarity in accordance to the literature [S2, S7].

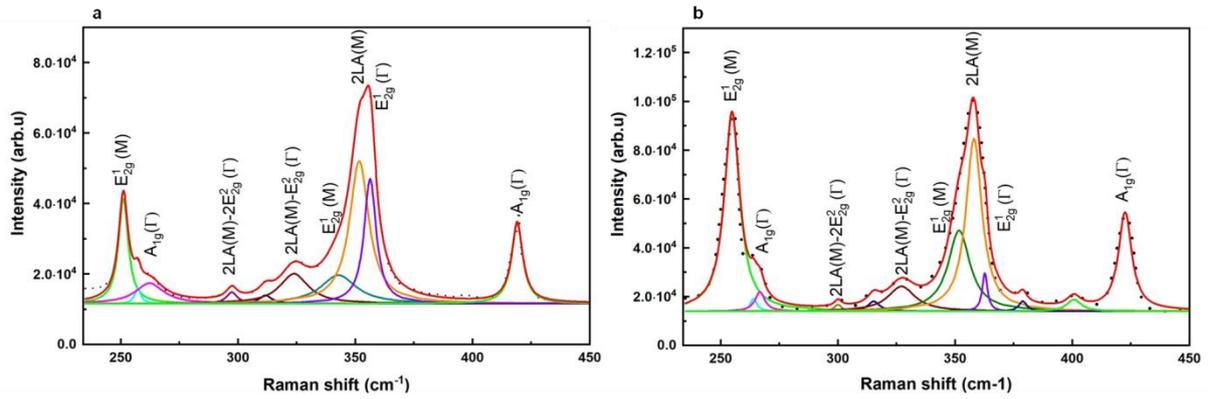

**Figure SI.6 | Raman spectra for the WS₂/WSe₂ HBL at room temperature (a) before and after (b) annealing.**

Here, changes in vibrational mode energies due to heterostructure assembly, such as a blue shift detected in the WS₂ $E_{2g}^1$ mode around 360 cm⁻¹ after annealing, are attributed to tension release. Moreover, the blue shift of the WS₂ $A_{1g}$ mode towards 450 cm⁻¹ might be a result of unintentional doping by adsorption of $O_2$ or $H_2O$, as previously reported [S8, S9]. As the WS₂ ML tends to release tension and shrink by adhesion to the deposited WSe₂ ML flake, compression strain is transferred to the capping ML of WSe₂ as a consequence. Based upon such compression, a blue shift in Raman modes of WSe₂ was probed in the literature [S10].

## S5. Excitons' μPL polarization degree for circularly-polarized excitation

To verify the excitonic character of the features in analogy to [S11], measurements with circular-polarized light with excitation–detection configurations σ⁺→ σ⁺ and σ⁺→ σ⁻, respectively, have been performed under CW excitation (at 1 mW). **Figure SI.7** shows PL measurements with different polarization configurations as blue and red spectra, respectively. This sheds light on the circular dichroism over the relevant spectral range. Co- and counter-polarization are established by polarization optics in the excitation as well as detection path in analogy to Ref. [S11]. As expected for excitonic species, which have been previously examined [S11], they exhibit a considerable degree of circular polarization, whereas defect states don't (cf. Refs. [S11, 48]).

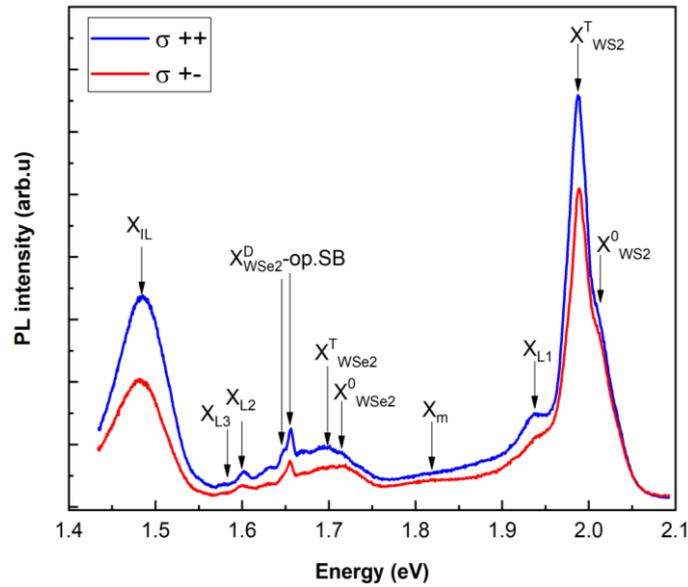

**Figure SI.7 | μPL spectra of co- and cross-polarized signal for the HBL at 10 K. (a), with features labeled as in Table 1.**

## S-References

S1.   O. Mey, F. Wall, L. M. Schneider, D. Günder, F. Walla, A. Soltani, H. Roskos, N. Yao, P. Qing, W. Fang, and A. Rahimi-Iman, "Enhancement of the Monolayer Tungsten Disulfide Exciton Photoluminescence with a Two-Dimensional Material/Air/Gallium Phosphide In-Plane Microcavity," ACS Nano 13, 5259–5267 (2019).

S2.   D. Kumar, B. Singh, P. Kumar, V. Balakrishnan, and P. Kumar, "Thermal expansion coefficient and phonon dynamics in coexisting allotropes of monolayer WS2 probed by Raman scattering," J. Phys. Condens. Matter 31, 505403 (2019).

S3.   Y. P. Varshni, "Temperature dependence of the energy gap in semiconductors," Physica 34, 149–154 (1967).

S4.   D. Christiansen, M. Selig, G. Berghäuser, R. Schmidt, I. Niehues, R. Schneider, A. Arora, S. M. De Vasconcellos, R. Bratschitsch, E. Malic, and A. Knorr, "Phonon sidebands in monolayer transition metal dichalcogenides," Phys. Rev. Lett. 119(18), 187402 (2017).

S5.   M. Selig, G. Berghäuser, A. Raja, P. Nagler, C. Schüller, T. F. Heinz, T. Korn, A. Chernikov, E. Malic, and A. Knorr, "Excitonic linewidth and coherence lifetime in monolayer transition metal dichalcogenides," Nat. Commun. 7(1), 13279 (2016).

S6   C. M. Chow, H. Yu, A. M. Jones, J. R. Schaibley, M. Koehler, D. G. Mandrus, R. Merlin, W. Yao, and X. Xu, "Phonon-assisted oscillatory exciton dynamics in monolayer MoSe2," npj 2D Mater. Appl. 1(1), 33 (2017).

S7.   L. M. Schneider, J. Kuhnert, S. Schmitt, W. Heimbrodt, U. Huttner, L. Meckbach, T. Stroucken, S. W. Koch, S. Fu, X. Wang, K. Kang, E. H. Yang, and A. Rahimi-Iman, "Spin-Layer and Spin-Valley Locking in CVD-Grown AA'- and AB-Stacked Tungsten-Disulfide Bilayers," J. Phys. Chem. C 123, 21813–21821 (2019).

S8.    L. Wang, X. Ji, F. Chen, and Q. Zhang, "Temperature-dependent properties of monolayer MoS2 annealed in an Ar diluted S atmosphere: An experimental and first-principles study," J. Mater. Chem. C 5, 11138–11143 (2017).

S9.    J. T. Mlack, P. Masih Das, G. Danda, Y.-C. Chou, C. H. Naylor, Z. Lin, N. P. López, T. Zhang, M. Terrones, A. T. C. Johnson, and M. Drndić, "Transfer of monolayer TMD WS2 and Raman study of substrate effects," Sci. Rep. 7, 43037 (2017).

S10.    O. Frank, G. Tsoukleri, J. Parthenios, K. Papagelis, I. Riaz, R. Jalil, K. S. Novoselov, and C. Galiotis, "Compression Behavior of Single-Layer Graphenes," ACS Nano 4, 3131–3138 (2010).

S11.    Y. You, X. Zhang, T. C. Berkelbach, M. S. Hybertsen, D. R. Reichman, and T. F. Heinz "Observation of biexcitons in monolayer WSe2," Nat. Phys. 11, 477–82 (2015).